\documentclass[british]{article}

\usepackage{amsmath,amssymb,mathrsfs,bm,hyperref}
\usepackage[all]{xy}

\newcommand{\Rho}{\mathrm{P}}
\newcommand{\id}{\mathrm{id}}
\newcommand{\zotimes}{\otimes_{\scriptscriptstyle Z}}
\newcommand{\ztimes}{\times_{\!\scriptscriptstyle Z}}
\newcommand{\smallZ}{\scriptscriptstyle Z}

\newtheorem{proposition}{Proposition}
\newtheorem{lemma}{Lemma}
\newcommand{\qed}{\hfill\(\square\)}

\title{Twisted Covariance as a Non  Invariant Restriction of the Fully Covariant DFR Model}

\author{Gherardo Piacitelli\thanks{SISSA, Via Beirut 2--4, 34151, Trieste, Italy. {\tt gherardo@piacitelli.org}}}

\begin{document}

\maketitle

\begin{abstract}
We discuss twisted covariance over the noncommutative 
spacetime algebra generated by the relations \([q_\theta^\mu,q_\theta^\nu]=i\theta^{\mu\nu}\),
where the matrix \(\theta\) is treated as fixed (not a tensor), and we refrain
from using the asymptotic Moyal expansion of the twists. 

We show that the tensor nature
of~\(\theta\) is only hidden in the formalism: in particular if~\(\theta\) 
fulfils the DFR conditions, the twisted Lorentz covariant model of the 
flat quantum spacetime may be
equivalently described in terms of the DFR model, if we agree to discard a huge
non invariant set of localisation states; it is only this last step which, if taken
as a basic assumption, severely breaks the relativity principle.

We also will show that the above mentioned, relativity breaking, 
{\itshape ad hoc} rejection of localisation states is an independent, unnecessary assumption, 
as far as some popular approaches to quantum field theory on the quantum Minkowski 
spacetime are concerned.

The above should raise some concerns about speculations on possible observable consequences of
arbitrary choices of \(\theta\) in arbitrarily selected privileged frames.

\end{abstract}

\tableofcontents

\section{Introduction}

There is nowadays some hope that noncommutative generalisations of geometry might
wake us up from the ultraviolet nightmare, 
and even open the way to a sound theory of quantum gravity. Several approaches are currently 
investigated; here we focus on a particular class of simplified models of a flat,
quantised spacetime.

We consider the (strong form of the) commutation relations 
\begin{equation}
\label{CR}
[q_\theta^\mu,q_\theta^\nu]=i\theta^{\mu\nu}
\end{equation}
among the selfadjoint spacetime coordinates \(q^0_\theta,q^1_\theta,q^2_\theta,q^3_\theta\), 
for some real, non degenerate, antisymmetric matrix~\(\theta\). In this paper, 
we adopt ``natural'' units: the light speed, the rationalised Planck constant
and the Planck length all are~1.

The above relations are understood as a quantisation of the \( 4 \)-dimensional 
Minkowski space-time. Interest in (a more general version of) these relations was  initially 
fueled by \cite{dfr}, where two Lorentz invariant conditions were imposed on the admissible
matrices~\(\theta\); the DFR conditions were deduced from a stability principle for the quantised
spacetime under localisation. See the original paper, or 
the less technical \cite{dfr_simple,karpaz}; and \cite{karpaz,corfu} for an outlook. Note that the DFR model ``fits comfortably into the deformation  
quantisation framework developed in \cite{rieffel}'', \cite{rieffel_96}.

Here, we fix a~\(\theta\) once and for all, 
fulfilling the DFR conditions (some comments on more general choices at the end of this
introduction). Together with~\(\theta\),
we consider its orbit \(\Sigma=\{\varLambda\theta\varLambda^t:\varLambda\in\mathscr L\}\) 
under Lorentz transformations, which is precisely
the family of all antisymmetric matrices fulfilling the DFR conditions. As a rule of thumb,
\(\theta,\theta'=\varLambda\theta\varLambda^t\in\Sigma\) will denote our fixed choice of a 
matrix in \(\Sigma\) 
and its Lorentz transform, and \(\sigma,\sigma'\in\Sigma\) will denote the dummy 
variable and its Lorentz transform.

The {\itshape ansatz} \eqref{CR} gives rise to the distinct models described here below.
\begin{itemize}
\item[(i)] \(\theta\) is fixed relatively to a particular classical observer in his own Lorentz 
frame (the `privileged' observer), 
and \eqref{CR} are the relations among the quantum coordinates driving Planck scale
phenomena in that frame; \(\theta\) transforms as a tensor.
The algebra of commutative functions is replaced with the algebra \(\mathcal K\) of 
compact operators; Weyl quantisation of classical symbols is defined in each Lorentz frame 
(connected with the privileged frame by \((\varLambda,a)\in\mathscr P\)) with
respect to \(\theta'=\varLambda\theta\varLambda^t\); correspondingly, in that frame the
Weyl calculus induces a twisted product \(\star_{\theta'}\).
All equations are Poincar\'e form-covariant,
but the relativity principle is broken at a fundamental level, since it is possible to classify
the observers accordingly to the \(\theta'\) they observe; such a classification is absolute 
with respect to the privileged\footnote{Of course the privilege is conventional and 
any other Lorentz frame with its corresponding commutation relations might play this role; `reference frame' would
be more appropriate, but would be confusing for evident reasons.} frame. We will call this 
model the {\bfseries reduced DFR model}, for reasons which will be clarified here below.
\item[(ii)] \(\mathcal C_0(\mathbb R^4)\) is replaced by \(\mathcal K\) as in the preceding case, but~\(\theta\) is kept constant in all frames, and the same twisted product \(\star_\theta\)
is used in all Lorentz frames. Ordinary Poincar\'e covariance is broken 
(at the level of formalism), but can be restored  in a twisted sense 
\cite{ch_tu,wess,aw}, 
using techniques from the theory of quantum groups \cite{drin,resh}.
In particular, with \(m(f\otimes g)=fg\) the ordinary pointwise
product of classical symbols, the twisted product may be written as
\(f\star_\theta g=m(F_\theta f\otimes g)\) for a suitable 
invertible operator \(F_\theta\) \cite{oeckl}, and Poincar\'e 
action is deformed in the coproduct, 
namely the ordinary action \(\gamma^{(2)}(L)f\otimes g=f'\otimes g'\) is deformed into
\(\gamma^{(2)}_\theta(L)={F_\theta}^{-1}\gamma^{(2)}F_\theta\); here \(f'(x)=f(L^{-1}x)\).
We will refer to this model as to the {\bfseries twisted covariant model}.
\item[(iii)] The matrices \(\sigma\in\Sigma\) 
label all possible equivalence classes of irreducible representations 
\([q_\sigma^\mu,q_\sigma^\nu]=i\sigma^{\mu\nu}\) of more general (DFR)
covariant commutation relations, so that the relations \eqref{CR} are not attached to a 
particular frame; all other representations are given by \(q_\sigma=\varLambda q_\theta\) if
\(\sigma=\varLambda\theta\varLambda^t\), and are equally important. 
The fully covariant represented 
coordinates can be obtained by direct integral techniques; they are related to the representation
of a trivial continuous field \(\mathcal E\) of C*-algebras over \(\Sigma\), 
where the Poincar\'e group acts by automorphisms. 
It is called the {\bfseries DFR model} \cite{dfr}.
\end{itemize}

In section \ref{main_sec:twisted} we will show that the twisted covariant 
model and the reduced DFR model are equivalent,
and that~\(\theta\) must be thought of as a tensor.  
Indeed, the twisted Poincar\'e
action maps the tensor product \(f\otimes g\) of symbols 
to \({F_\theta}^{-1}(F_\theta f\otimes g)'=
{F_\theta}^{-1}F_{\theta'}f'\otimes g'\), where primes indicate ordinary Poincar\'e action.
It follows that the \(\theta\)-twisted product of the twisted transformation of \(f\otimes g\)
is precisely the same as the \( \theta'\)-twisted product of the untwisted
transformation of \(f\otimes g\), namely
\begin{equation}
m_\theta(\gamma^{(2)}_\theta(L)f\otimes g)=
m(F_\theta {F_\theta}^{-1}F_{\theta'}f'\otimes g')=m(F_{\theta'}f'\otimes g')=f'\star_{\theta'}g';
\end{equation}
so that twisted covariance is formally equivalent to undeformed 
covariance\footnote{The transformation rule \eqref{lasannotutti} 
for twisted products 
was first established in this context in \cite[eq. (4.5)]{dfr}, in
momentum space. It first appeared as an equation in position space
in \cite{rieffel} and, in the case of more general linear affine spacetime 
transformations, in \cite{gb}.}
\begin{equation}\label{lasannotutti}
(f\star_\theta g)'=f'\star_{\theta'}g',
\end{equation}
if~\(\theta\) is treated as a tensor.  Hence keeping~\(\theta\) constant in all frames,
while twisting the coproduct, is equivalent to treating~\(\theta\) as a tensor, while
keeping the ordinary 
(undeformed) coproduct. To embody this purely formal comment with a meaningful
interpretation, we will {\itshape deduce} from twisted covariance and 
Weyl quantisation that, even agreeing to formally treat~\(\theta\) as a constant matrix,
the commutation relations among the coordinates~---~as they are seen by an unprivileged
observer~---~do transform as a tensor. To put it in another way, 
twisted covariance itself is incompatible with performing
the Weyl quantisation in all frames with the same coordinates \eqref{CR}.

Moreover, in section \ref{main_sec:tough} 
we will show that the reduced DFR model can be obtained from the full 
DFR model up to rejecting a huge, non invariant 
class of otherwise admissible localisation states (states on \(\mathcal E\)). Precisely,
only the states which are pure on the centre of \(\mathcal E\) and concentrated on~\(\theta\) 
are available to the privileged observer; and these states are mapped by the dual action
of the Poincar\'e group precisely to the localisation
states which only are available to the observer 
in the correspondingly transformed frame. 
This criterion for rejecting otherwise admissible DFR localisation states will
be called here~\(\theta\)-universality.

This will lead us in the conclusions 
to formulate a natural criticism, which can be
summarised in the following question: since a fully covariant model is available, which 
reproduces the twisted covariance formalism at the price of an additional independent
assumption which breaks the relativity principle, why should we make that assumption?
These results and the criticism were already anticipated in \cite{piac}. 
To strengthen our criticism, we 
will show in section \ref{sec_main:fields} 
that~\(\theta\)-universality does not play any crucial role
in some recent approaches to quantum field theory. In particular, the approach
of \cite{lg_1,buchholz,gandalf-harald} on one side has no relations with
\(\theta\)-universality (as the authors themselves are well aware of); on the other side,
it provides a formalism  which easily allows for showing that the so called 
``twisted CCR'' (\cite{bal_1,bal_2}), although developed within \(\theta\)-universality, 
do not critically rely on it, and could be understood fibrewise over \(\Sigma\). 
Of course, the above results entail a fundamental objection against speculations on possible
observable consequences of~\(\theta\)-universality within this particular class of models.

We also will provide some clarifications on the issue of coordinates of many events
in appendix \ref{app:many_var}, and some remarks on Wightman functions; 
as a side comment, 
we will prove that the braided commutation
relations among the coordinates of many events, introduced in \cite{fw}, only have
trivial regular representations.

\bigskip 
\centerline{***}
\bigskip
We close this introduction with a few remarks. 
The discussion of twisted covariance and the proof that~\(\theta\) is a tensor 
does not rely on~\(\theta\) fulfilling the DFR conditions, which we only required on the purpose
of making contact with the DFR model. 
Although here the explicit functional form of the integral 
kernels is given for an invertible~\(\theta\) (as DFR matrices are),
the formalism can easily be generalised (see e.g.\ \cite{rieffel,gb}) 
to the case of a non invertible matrix, including
the case of time-space commutativity\footnote{Note however that time-space commutativity is
not preserved by Lorentz transformations.}. The assumption that dimension of spacetime 
is 4 also is not necessary.

There is, however, a more subtle implicit assumption: for the symbolic calculus to be a
faithful
replacement of the full C*-algebra arising from Weyl quantisation, 
irreducible representations of the commutation relations 
should exist and be unique. 
By adapting the argument of \cite{dfr}, this certainly is the case
whenever the degeneracy space of \(\theta\) has even codimension, in which case we can
rely on von Neumann theorem \cite{vn}. 
If, otherwise, the existence of representations is not known, one should keep in mind the quantum
replacement of a well known principle: 
``no deformation without representation!''

\section{Twisted Covariance}
\label{main_sec:twisted}

Here, we will carefully describe the twisted covariant model, using integral kernels
in position and momentum space instead of the Moyal expansion. Then we will
show that the twisted covariant model is equivalent to the reduced DFR model at a formal
level, and we will give evidence that the tensor nature of \(\theta\) is enforced by the
interpretation.

\subsection{Weyl Quantisation and Twisted Products}\label{sec:tp}
When integrated in their Weyl form
\[
e^{ih_\mu q_\theta ^\mu}e^{ik_\mu q_\theta^\mu}=e^{-\frac{i}{2}h_\mu\theta^{\mu\nu} k_\nu}e^{i(h+k)_\mu q_\theta ^\mu},
\] 
the relations \eqref{CR} induce a symbol calculus
through Weyl quantisation \(W_\theta(f)=\int\check f(k)e^{ik_\mu q_\theta ^\mu}\) and the corresponding
twisted product \({\star_\theta}\) \cite{weyl}, so that
\[
W_\theta(f)W_\theta(g)=W_\theta(f{\star_\theta} g), \quad W_\theta(\bar f)=W_\theta(f)^*.
\]
Weyl quantisation is defined on \(L^1\cap\widehat{L^1}\), though in principle it could
be extended to a much wider class of distributions by bitransposition.\footnote{S.\ Doplicher,
private conversation.} From now on, we will systematically use the shorthand
\[
L^1=L^1(\mathbb R^{4})
\]
with respect to the translation invariant Lebesgue measure.
It is usually more convenient
to work in momentum space,\footnote{\label{fourier_conventions}
We agree on the following, asymmetric conventions:
\[
\check f(y)=\frac{1}{(2\pi)^{4}}\int\limits_{\mathbb R^{4}} dx\;f(x)e^{-ix_\mu y^\mu},
\quad\quad
\hat f(y)=\int\limits_{\mathbb R^{4}} dx\;f(x)e^{ix_\mu y^\mu}.
\]
} where the twisted product becomes a twisted convolution product \cite{vn}:
\[
f{\star_\theta} g=\widehat{\check f{\times_\theta} \check g};
\]
standard computations yield\footnote{In the context of canonical quantisation,
the use of twisted products was first 
advocated in \cite{weyl}; the first explicit definition was given in momentum
space in \cite{vn}. Here we strictly adhere to the spirit of those 
authors, where (Weyl) quantisation is the primary concept, and twisted
products only are interesting as ancillary tools, 
in that they provide a symbolic calculus for the operators resulting from
quantisation. This is different from the spirit of many followers of
the influential paper \cite{bayen}; indeed, they take 
twisted products as the fundamental objects of a quantisation, 
without making explicit  the C*-algebraic content. For related 
developments in this direction, see e.g.\ 
\cite{maillard,michor}.}
\begin{align*}
&(\varphi{\times_\theta}\psi)(k)=\int dh\varphi(h)\psi(k-h)e^{-\frac{i}{2}h\theta k}, \quad\varphi,\psi\in
L^1,\\
&(f{\star_\theta} g)(x)=
\frac{2}{(2\pi)^{4}|\det\theta|}
\iint du\,dv f(x+u)g(x+v)e^{2iu\theta^{-1}v},
\quad f,g\in L^1\cap\widehat{L^1},
\end{align*}
where from now on we use the shorthands \(hx=h_\mu x^\mu\), \(h\theta k=h_\mu\theta^{\mu\nu}k_\nu\),
and so on.\footnote{\label{foot:matrix}In matrix notation, 
\(h_\mu\theta^{\mu\nu}k_\nu=h^tG\theta Gk\) with \(h,k\) column vectors, where
\(\varLambda\) by definition fulfils \(\varLambda^t G\varLambda=G\) and the metric
matrix 
\(G=(g^{\mu\nu})=(g_{\mu\nu})=\text{diag}(1,-1,-1,-1)\) fulfils \(G^2=I\);
in particular we have \(\varLambda^{-1}=G\varLambda^tG\). Here the superscript \(t\)
denotes rows-by-columns transposition: \({(\varLambda^t)^\mu}_\nu={\varLambda^\nu}_\mu\).}
\(L^1\)~equipped with the twisted convolution product and the involution 
\(\varphi^*(k)=\overline{\varphi(-k)}\) is a Banach *-algebra \(\mathcal E^0_\theta\)
of which \(\pi_\theta(\varphi)=\int\varphi(k)e^{ikq_\theta }\) 
is a *-representation. Of course, \(\pi_\theta(\check f)=W_\theta(f)\). The universal enveloping C*algebra of \(\mathcal E^0_\theta\) is the algebra \(\mathcal K\),
of compact operators on the separable, infinite dimensional Hilbert space.

\subsection{Drinfel'd Twists}\label{sec:dt}

The twisted product has been recognised by Oeckl \cite{oeckl} (see also the 
earlier \cite{watts}) 
as a twist in the sense of \cite{drin,resh} 
(see \cite{aschieri_ln} for a review). In turn, this kind of deformations
are examples of the more general construction discussed 
in  \cite[Theorem 8]{gerst}.

Let us restrict
ourselves to the functions in the Schwartz space, which is naturally
recognised as a subspace of 
\(L^1\cap\widehat{L^1}\), and where the Fourier transform acts continuously and invertibly. Let
\[
{\mathscr S}\subset\bigoplus_{n=1}^\infty\mathscr S^{(n)}
\] 
be the space of sequences \(\{f_n\in\mathscr S^{(n)}\}\) with \(f_n\equiv 0\) eventually, 
where  we write \(\mathscr S^{(n)}\) for the Schwartz space
on \(\mathbb R^{4n }\). In what follows we will implicitly identify
\(\mathscr S^{(n)}\otimes \mathscr S^{(m)}=\mathscr S^{(n+m)}\).

If
\[
m^{(2)}:\mathscr S^{(2)}\rightarrow\mathscr S^{(1)}
\]
is the diagonal map
\[
(m^{(2)}\xi)(x)=\xi(x,x),
\]
then \(fg=m^{(2)}(f\otimes g)\) is the ordinary pointwise multiplication, and
\[
f{\star_\theta} g=m^{(2)}(F_\theta f\otimes g)),
\]
where the map \(F^{(2)}_\theta:\mathscr S^{(2)}\rightarrow\mathscr S^{(2)}\)
is defined by
\[
(F^{(2)}_\theta\xi)(x,y)=
\frac{2}{(2\pi)^{4}|\det\theta|}
\iint du\,dv \xi(x+u,y+v)e^{2iu\theta^{-1}v},\quad \xi\in\mathscr S^{(2)}.
\]
Note that  \(F^{(2)}_\theta\) is not uniquely defined by the above requirement, since whatever
other choice agreeing 
on the diagonal set \(\{x=y\}\) would do the 
required job. Here we always will refer to the above choice.

If \(f,g\) in addition are entire analytic, then
\begin{equation}\label{eq:moyal_twist}
F_\sigma f\otimes g=
m\left(e^{\frac{i}{2}\theta^{\mu\nu}\partial_\mu\otimes\partial_\nu}f\otimes g\right),
\end{equation}
which is a compact notation for the Moyal expansion
\begin{align*}
\mathscr M[f\star_\theta g](x)=&f(x)g(x)+\\&+\sum_{n=1}^N\frac{(i/2)^n}{n!}
\theta^{\mu_1\nu_1}\dotsm \theta^{\mu_n\nu_n}(\partial_{\mu_1}\dotsm \partial_{\mu_n} f)(x)
(\partial_{\nu_1}\dotsm \partial_{\nu_n} g)(x);
\end{align*}
see \cite{g_parigi} for some comments on the drawbacks of this notation in this context; and 
\cite{egv} for a thorough discussion of the analytic subtleties (or the more 
recent \cite{soloviev}). 

More generally if, \((m^{(n)}\xi)(x)=\xi(x,x,\dotsc,x)\), \(\xi\in\mathscr S^{(n)}\),
then
\[
f_1{\star_\theta} f_2{\star_\theta}\dotsm {\star_\theta}f_n=
m^{(n)}(F_\theta^{(n)}f_1\otimes f_2\otimes\dotsm\otimes f_n),
\]
where the explicit action of \(F_\theta^{(n)}\) can be obtained from the 
kernels computed in \cite[appendix C]{dfr}.

Equivalently in momentum space, with 
\(c^{(n)}(\varphi_1\otimes\dotsm\otimes\varphi_n)=\varphi_1\times\dotsm\times\varphi_n\) 
the ordinary convolution product, one finds
\[
\varphi_1{\times_\theta} \varphi_2{\times_\theta}\dotsm {\times_\theta}\varphi_n=c^{(n)}(T^{(n)}_\theta\varphi_1\otimes\varphi_2\otimes\dotsm\otimes\varphi_n),
\]
where the operator
\[
(T^{(n)}_\theta\xi)(k_1,\dotsc,k_n)=e^{-\frac{i}{2}\sum_{i<j}k_i\theta k_j}\xi(k_1,\dotsc,k_n), 
\quad \xi\in\mathscr S^{(n)},
\]
is evidently invertible with inverse 
\[
{T^{(n)}_\theta}^{-1}=T^{(n)}_{-\theta}.
\]
From this and invertibility of Fourier transform 
it follows that each \(F^{(n)}_\theta\) also is invertible with
\[
{F^{(n)}_\theta}^{-1}=F^{(n)}_{-\theta}.
\]
By construction the diagram
\[
\xymatrix{
{\mathscr S}\ar@<1ex>[r]^{{F_\theta}}\ar@<1ex>[d]^{\check{}}&{\mathscr S}\ar@<1ex>[l]^{{F_{-\theta}}}\ar[r]^{ m}\ar@<1ex>[d]^{\check{}}
&\mathscr S^{(1)}\ar@<1ex>[d]^{\check{}}\\
{\mathscr S}\ar@<1ex>[u]_{\hat{}}\ar@<1ex>[r]^{{T_\theta}}&{\mathscr S}\ar@<1ex>[l]^{{T_{-\theta}}}
\ar@<1ex>[u]_{\hat{}}\ar[r]^{c}&\mathscr S^{(1)}\ar@<1ex>[u]^{\hat{}}
}
\]
is commutative, where \(m(\{f_n\})=\sum_nm^{(n)}(f_n)\), \({F_\theta}=\bigoplus_nF_\theta^{(n)}\), 
and analogously for the other maps.

As for explicit formulae, it is well known that
\[
c^{(n)}(\varphi)(k_n)=\int\!\dotsm\!\int 
dk_1\dotsm dk_{n-1}\,\varphi\left(k_1,\dotsm,k_{n-1},k_n-\sum_i^{n-1}k_i\right).
\]
The proof by induction that 
\begin{align*}
c_\theta^{(n)}(\varphi)(k_n)&=c^{(n)}(T^{(n)}_\theta\varphi)(k_n)=\\
&=\int\!\dotsm\!\int dk_1\dotsm dk_{n-1}\,\varphi\left(k_1,\dotsm,k_{n-1},k_n-\sum_i^{n-1}k_i\right)
e^{-\frac{i}{2}\sum_{i<j}k_i\theta k_j}
\end{align*}
is the solution of the recursive equation
\[
c_\theta^{(n+1)}=c_\theta^{(2)}\circ(\id \otimes c_\theta^{(n)})
\]
is a routine computation \cite{dfr}.

\subsection{Twisting the Action of Lorentz Transformations}\label{sec:twist_1}

There is an action \(\gamma^{(n)}\) of the full
Poincar\'e group \(\mathscr P\) by endomorphisms 
on \((\mathscr S^{(n)},\cdot)\), given by
\[
(\gamma^{(n)}(L)f)(x)=(\det\varLambda)^nf(L^{-1}x_1,\dotsc,L^{-1}x_n),
\quad L=(\varLambda,a)\in\mathscr P.
\]
which is such that \(m^{(n)}\circ\gamma^{(n)}(L)=
\gamma^{(1)}(L)\circ m^{(n)}\). Equivalently in momentum space there is an action
\[
(\beta^{(n)}(L)\varphi)(k_1,\dotsc,k_n)=(\det\varLambda)^n
e^{-ia\sum_jp_j}\varphi(\varLambda^{-1}k_1,\dotsc,\varLambda^{-1}k_1),
\]
so that the diagram
\[
\xymatrix{
\mathscr S\ar[d]^c&{\mathscr S}\ar[l]_{\beta(L)}\ar[d]^c\ar[r]^{\hat{}}&{\mathscr S}\ar[r]^{\gamma(L)}\ar[d]_m&\mathscr S\ar[d]_m\\
\mathscr S^{(1)}&{\mathscr S}^{(1)}\ar[l]_{\beta^{(1)}(L)}\ar[r]^{\hat{}}&{\mathscr S}^{(1)}\ar[r]^{\gamma^{(1)}(L)}&\mathscr S^{(1)}
}
\]
is commutative, where all horizontal arrows are invertible.

According to \cite{ch_tu,wess,aw}, one may wish to look for 
a deformed action \(\gamma_\theta\) (\(\beta_\theta\) in momentum space) of the
Poincar\'e group on \(\mathscr S\) which is ``compatible with the twisted algebraic structure'',
namely such that the diagram
\begin{equation}\label{magari}
\xymatrix{
\mathscr S\ar[d]^{c_\theta}&{\mathscr S}\ar[l]_{\beta_\theta(L)}\ar[d]^{c_\theta}\ar[r]^{\hat{}}&{\mathscr S}\ar[r]^{\gamma_\theta(L)}\ar[d]_{m_\theta}&\mathscr S\ar[d]_{m_\theta}\\
\mathscr S^{(1)}&{\mathscr S}^{(1)}\ar[l]_{\beta_\theta^{(1)}(L)}\ar[r]^{\hat{}}&{\mathscr S}^{(1)}\ar[r]^{\gamma_\theta^{(1)}(L)}&\mathscr S^{(1)}
}
\end{equation}
is commutative, where \(m_\theta=m\circ F_\theta\), \(c_\theta=c\circ T_\theta\), and again
horizontal arrows are invertible.

This can be achieved by taking
\[
\gamma_\theta(L)=F_{-\theta}  \gamma(L)  F_{\theta},\quad n>1
\]
or, in momentum space,
\[
\beta_\theta(L)=T_{-\theta}  \beta(L)  T_{\theta},\quad n>1;
\]
note that the action on \(\mathscr S^{(1)}\) is unchanged:
\begin{equation}\label{1_the_same}
\gamma_\theta^{(1)}(L)=\gamma^{(1)}(L), \quad\beta_\theta^{(1)}(L)=\beta^{(1)}(L).
\end{equation}
It is self evident that 
\[\gamma_\theta(L)\gamma_\theta(L')=\gamma_\theta(LL'),\quad
\gamma_\theta(I)=\id ,
\] 
so that we have an action of \(\mathscr P\)
on \(\mathscr S\), indeed. Moreover, a straightforward computation shows that
\[
m_\theta\circ \gamma_\theta(L)=\gamma_\theta^{(1)}(L)\circ m_\theta,
\]
which proves that the diagram \eqref{magari} is commutative, as desired.

Equivalence of the above with the formalism developed in
\cite{ch_tu,wess,aw} is confirmed  by the following 
\begin{proposition}
For \(\varepsilon\in\mathbb R\), let 
\(\varLambda(\varepsilon)=({\varLambda(\varepsilon)^\mu}_\nu)=
({g^\mu}_\nu+\varepsilon{\omega^\mu}_\nu)+o(\varepsilon)\)
be a proper orthochronous Lorentz transformation, where
\[
{\omega^\mu}_\nu=-{\omega_\nu}^\mu, \quad 
\varLambda(\varepsilon)^{-1}=\varLambda(-\varepsilon)+o(\varepsilon)
\]  
and \(g=(g^{\mu\nu})\) is the Lorentz metric.

Moreover, let \(\kappa^\mu\) denote the operator of multiplication
\((\kappa^\mu \varphi)(k)=k^\mu \varphi(k)\), and 
\((\partial_\mu\varphi)(k)=\partial\varphi/\partial k^\mu\).

Finally, with \(X\) a continuous linear operator on \(\mathscr S^{(1)}\),
we define \(\Delta[X]=X\otimes I+I\otimes X\). 

Then
\begin{align*}
\left.\frac{d}{d\varepsilon}\beta^{(1)}((\varLambda(\varepsilon),0))\right|_{\varepsilon=0}&=
-{\omega^\mu}_\nu\kappa^\nu\partial_\mu,\\   
\left.\frac{d}{d\varepsilon}\beta^{(2)}((\varLambda(\varepsilon),0))\right|_{\varepsilon=0}
&=\Delta[-{\omega^\mu}_\nu\kappa^\nu\partial_\mu],\\
\left.\frac{d}{d\varepsilon}\beta^{(2)}_\theta(\varLambda(\varepsilon),0))\right|_{\varepsilon=0}&=
\Delta_\theta[-{\omega^\mu}_\nu\kappa^\nu\partial_\mu]=\\
&=\Delta[-{\omega^\mu}_\nu\kappa^\nu\partial_\mu]+
\frac{i}{2}({\omega^\mu}_\rho\theta_{\mu\sigma}+
{\omega^\nu}_\sigma\theta_{\rho\nu})\kappa^\rho\otimes\kappa^\sigma.
\end{align*}
where
\[
\Delta_\theta[X]=T_{-\theta}^{(2)} \Delta[X]T_{\theta}^{(2)}.
\]

Moreover,
\[
(\Delta_\theta\otimes\id)\circ\Delta_\theta[-{\omega^\mu}_\nu\kappa^\nu\partial_\mu]=
(\id\otimes\Delta_\theta)\circ\Delta_\theta[-{\omega^\mu}_\nu\kappa^\nu\partial_\mu].
\]
\end{proposition}

The proof consists of straightforward computations 
which we refrain from spelling;
when applied to the generators of infinitesimal Lorentz
transformations,
the map \(\Delta\)  may be recognised as the
(represented action of the) coproduct of 
primitive elements in the universal enveloping Lie algebra
of the Lorentz group; the last statement in the proposition
is a check of coassociativity
on primitive elements. See e.g.\ \cite{aschieri_ln} for a short and readable 
introduction to the language of Hopf algebras and twists, and to its 
applications to twisted covariance.

\subsection{Strict Covariance of the Commutation Relations}\label{sec:wow}

We now turn to the interpretation of twisted covariance. We have seen that the formalism 
of twisted covariance allows all observers for using 
the same matrix~\(\theta\) to twist the product
in all Lorentz frames; this is commonly interpreted by saying that~\(\theta\) is a universal
invariant matrix which does not transform as a tensor. This view of course entails a fundamental
breakdown of the relativity principle.

However, already from the point of view of analytic expressions, the above view is certainly
not the only possible interpretation of the situation.

Let \(\varphi\in L^1(\mathbb R^{4n})\) and 
\(L=(\varLambda,a)\) be a Poincar\'e transformation. Recalling that
\[
(\beta^{(n)}(L)\varphi)(k_1,\dotsc,k_n)=(\det\varLambda)^ne^{-ia\sum_ik_i}
\varphi(\varLambda^{-1}k_1,\dotsc,\varLambda^{-1}k_n),
\]
and
\[
(T^{(n)}_\theta\varphi(k_1,\dotsc,k_n))=e^{-\frac{i}{2}\sum_ik_i\theta k_j}\varphi(k_1,\dotsc,k_n),
\]
it follows immediately that
\begin{equation}\label{eureka_2}
\beta^{(n)}(L)T_\theta^{(n)}=T_{\theta'}^{(n)}\beta^{(n)}(L),
\end{equation}
where
\[
\theta'=\varLambda\theta\varLambda^t
\]
or, in Einstein notation,
\[
{\theta'}^{\mu\nu}={\varLambda^\mu}_{\mu'}{\varLambda^\nu}_{\nu'}\theta^{\mu'\nu'}.
\]

As a consequence of \eqref{eureka_2}, the twisted action fulfils
\[
\beta^{(n)}_\theta(L)={T^{(n)}_{\theta}}^{-1}\beta^{(n)}(L)T^{(n)}_\theta=
{T^{(n)}_{\theta}}^{-1}T^{(n)}_{\theta'}\beta^{(n)}(L)
\]
It easily follows that
\begin{align*}
c^{(n)}_\theta(\beta^{(n)}_\theta(L)\varphi)&=c^{(n)}(T_\theta^{(n)}\beta^{(n)}_\theta(L))=\\
&=c^{(n)}({T_\theta^{(n)}}^{-1}T_\theta^{(n)}T_{\theta'}^{(n)}\beta^{(n)}\varphi)=\\
&=c^{(n)}(T_{\theta'}\beta^{(n)}(L)\varphi)=\\
&=c^{(n)}_{\theta'}(\beta^{(n)}(L)\varphi).
\end{align*}

Indeed, we proved the following
\begin{proposition}\label{lemma:bah}
Let \(f_i\in L^1\cap \widehat{L^1}\), \(i=1,2,\dotsc,n\), 
and \(L=(\varLambda,a)\in\mathscr P\). Then 
\begin{gather*}
m_\theta^{(n)}(\gamma^{(n)}_\theta(L)f_1\otimes f_2\otimes\dotsm\otimes f_n)=
f_1'\star_{\theta'}f_2'\star_{\theta'}\dotsm\star_{\theta'}f_n',\\
c_\theta^{(n)}(\beta^{(n)}_\theta(L)\check f_1\otimes \check f_2\otimes\dotsm\otimes \check f_n)=
\check f_1'\times_{\theta'}\check f_2'\times_{\theta'}\dotsm\times_{\theta'}\check f_n',
\end{gather*}
where 
\[
f_i'(x)=f_i(\varLambda^{-1}(x-a))
\] 
and
\[
{\theta'}^{\mu\nu}={\varLambda^\mu}_{\mu'}{\varLambda^\nu}_{\nu'}\theta^{\mu'\nu'}.
\]
\end{proposition}

As a consequence of this proposition, 
twisted covariance as expressed by diagram \eqref{magari}
is completely equivalent to
\[
(f_1\star_\theta f_2\star_\theta\dotsm\star_\theta f_n)'=
(f_1'\star_{\theta'} f_2'\star_{\theta'}\dotsm\star_{\theta'} f_n').
\]

In other words, twisted Lorentz covariance with invariant twisted products is mathematically
equivalent to ordinary Lorentz covariance with covariant twisted products. Thus, the 
statement that~\(\theta\) is constant
(not a tensor) in all frames is at least questionable. 

Although this alternative point of view might seem more appealing as it restores
formal covariance, this is not yet a sufficient reason to prefer it. 
Formal covariance only is meaningful
if one trusts the relativity principle, which in the present case is broken anyway by the choice
of a fixed~\(\theta\) in a given reference frame (we will comment on this later in this paper);
notwithstanding the covariant aspect of equations, still it would be 
possible to classify the observers according to the \(\theta'\) they see in their own frame.
As far as we accept to break the relativity principle, the two formalisms have the same
dignity.

In order to take a decision about
which view is more adherent to our purposes,
we must endow \(i\theta\) with its physical interpretation: 
it is the commutator of the quantum coordinates in a given frame; twisted products
only are an auxiliary device for computing products of Weyl--quantised functions.

Hence, the right question to ask is: which commutation rules does the primed observer observe?
In order to answer it, we assume that the quantum coordinates \(q'\) 
for the primed observer fulfil some {\itshape a priori} unknown commutation rules. Whatever these
commutation rules are, we assume that the primed observer adopts the 
Weyl quantisation
\[
W'(f)=\int dk\check f(k)e^{ikq'},\quad f\in L^1\cap\widehat{L^1};
\]
she also defines her own~---~{\itshape a priori} 
unknown~---~twisted product \(\star'\) by requiring that
\[
W'(f)W'(g)=W'(f\star'g),\quad f,g\in L^1\cap\widehat{L^1}.
\]

Now we are ready to use twisted covariance:
whatever the commutation relations among the \({q'}^\mu\)'s do appear to the new observer, 
the identity
\[
W'(m_\theta(\gamma^{(2)}_\theta(L)f\otimes g)=W'(f')W'(g')
\]
must hold true, where
\[
f'(x)=f(\varLambda^{-1}(x-a)),\quad g'(x)=g(\varLambda^{-1}(x-a)).
\] 
We compute
\begin{eqnarray*}
\lefteqn{W'(m^{(2)}_\theta(\gamma^{(2)}_\theta(L)f\otimes g))=
\int dk\;c^{(2)}(\beta^{(2)}(L)T^{(2)}_\theta\check f\otimes\check g)(k)e^{ik{q'}}=}\\
&=&\iint dhdk\; e^{ik({q'}-a)}e^{-\frac{i}{2}h(\varLambda\theta\varLambda^t)k)}
\check f(\varLambda^{-1}h)
\check g(\varLambda^{-1}(k-h))=\\
&=&\iint dhdk'\; e^{i(k'+h)({q'}-a)}e^{-\frac{i}{2}h(\varLambda\theta\varLambda^t)k'}\check f(\varLambda^{-1}h)
\check g(\varLambda^{-1}(k'))=\\
\lefteqn{=W'(f')W'(g')=\iint dhdk\;e^{-i(h+k)a}e^{ih{q'}}e^{ik{q'}}
\check f(\varLambda^{-1}h)\check g(\varLambda^{-1}k),}
\end{eqnarray*}
from which (using the arbitrariness of \(f,g\)) 
the Weyl relations for the \({q'}^\mu\)'s are immediately recovered:
\[
e^{ih{q'}}e^{ik{q'}}=e^{-\frac{i}{2}h\theta'k}e^{i(h+k)q'},
\]
which are the Weyl form of the relations
\begin{equation}\label{eureka_1}
[{q'}^\mu,{q'}^\nu]=i\theta'^{\mu\nu}.
\end{equation}
We found in the new reference frame
\begin{gather*}
W'=W_{\theta'},\\
{q'}^\mu=q^\mu_{\theta'}={\varLambda^\mu}_\nu q^\nu_{\theta},\\
\star'=\star_{\theta'}.
\end{gather*}

The tensor nature of~\(\theta\)
is thus established in the interpretation, too.

\section{From the DFR model to Twisted Covariance}\label{main_sec:tough}

In this section, we will derive the reduced DFR model (and thus, according to the
discussion of the previous section, the twisted covariant model) from
the fully covariant DFR model, making the additional, independent 
assumption of \(\theta\)-universality.

Essentially, we will show that \(\theta\)-universality is equivalent to the prescription
of projecting, in each Lorentz frame, the full spacetime algebra on its fibre over \(\theta'\), 
where \(\theta'\) is the Lorentz transform of the \(\theta\) corresponding to the
privileged frame.

\subsection{The DFR algebra}\label{sec:dfr_recall}

We begin by shortly recall some basic facts about the DFR algebra and its continuous
sections as a continuous field of C*-algebra. We also will take the 
opportunity of writing the full DFR twisted product in terms of a fibrewise Drinfel'd twist,
as a complementary indication that the formalism has a covariant reformulation.

Following closely \cite{dfr}, 
we equip the space \(\mathcal C_0(\Sigma,L^1)\) of the 
\(L^1\)-valued continuous functions \((\sigma\mapsto\varphi(\sigma;\cdot))\)
vanishing at infinity with the product (fibrewise twisted convolution)
\begin{equation}\label{eq:prod_dist}
(\varphi\ztimes\psi)(\sigma;\cdot)=\varphi(\sigma,\cdot)\times_\sigma\psi(\sigma;\cdot),
\end{equation}
the involution 
\[
\varphi^*(\sigma;k)=\overline{\varphi(\sigma;-k)},
\]
and the action
\[
(\beta((\varLambda,a))\varphi)(\sigma;k)=(\det\varLambda)
e^{-ika}\varphi(\varLambda^{-1}\sigma{\varLambda^{-1}}^t;\varLambda^{-1}k)
\]
of the Poincar\`e group. The norm
\[
\|\varphi\|_{0,1}=\sup_\sigma\|\varphi(\sigma;\cdot)\|_{L^1}
\]
makes it a Banach *-algebra which we denote by \(\mathcal E^{(0)}\).

According to \cite[Theorem 4.1]{dfr}, there exists a unique C*-norm \(\|\cdot\|\) on \(\mathcal E^{(0)}\), and the C*-completion \(\mathcal E\) of \(\mathcal E^{(0)}\) is isomorphic as
a continuous field of C*-algebras  
to the trivial continuous field \(\mathcal C_0(\Sigma,\mathcal K)\), where the
standard fibre \(\mathcal K\) is the algebra  of compact operators on the separable, 
infinite dimensional Hilbert space. Moreover, the action \(\beta\) extends by continuity
to an isomorphism \(\alpha:\mathscr P\rightarrow\text{aut}(\mathcal E)\).

In particular, for each \(\sigma\), one may form the algebra \(\mathcal E^{(0)}_\sigma\)
by restriction to \(\sigma\); namely as a Banach space \(\mathcal E^{(0)}_\sigma=L^1\);
the product is of course \(\times_\sigma\). For each
\(\sigma\) the unique C*-completion of \(\mathcal E^{(0)}_\sigma\) is \(\mathcal K\); the natural
 inclusions \(\mathcal E^{(0)}\subset\mathcal E\) and
\(\mathcal E^{(0)}_\sigma\subset\mathcal K\) will be implicitly understood.

The maps \(\Pi_\sigma:\mathcal E^{(0)}\mapsto \mathcal E^{(0)}_\sigma\) defined by
\[
(\Pi_\sigma\varphi)(\cdot)=\varphi(\sigma,\cdot),\quad\varphi\in\mathcal E^{(0)},
\]
extend
by continuity to *-homomorphisms  \(\Pi_\sigma:\mathcal E\mapsto \mathcal K\); they
must be understood as projections onto the fibre over \(\sigma\).

The fibrewise twisted convolution can be written in terms of a fibrewise Drinfel'd twist, too, if we
define the fibrewise tensor product of sections\footnote{By C*-completion, the
fibrewise tensor product extends to the tensor product of \(Z\)-moduli of two copies
of \(\mathcal E\), where \(Z\) is the centre of the multipliers algebra \(M(\mathcal E)\). 
This explains
the notation \(\zotimes\).}
\[
(\varphi\zotimes\psi)(\sigma;h,k)=\varphi(\sigma,h)\psi(\sigma,k).
\]
Then ordinary fibrewise convolution is 
\[
c^{(2)}(\varphi\zotimes\psi)(\sigma;k)=(\varphi\times\psi)(\sigma,k)
\]
and fibrewise twisted convolution is
\[
c^{(2)}_{\smallZ}(\varphi\zotimes\psi)(\sigma;k)=(\varphi\ztimes\psi)(\sigma,k).
\]
The twist operator now depends on \(\sigma\):
\[
(T_{\smallZ}^{(2)}\varphi\zotimes\psi)(\sigma;h,k)=e^{-\frac{i}{2}h\sigma k}(\varphi\zotimes\psi)(\sigma;h,k),
\]
and of course
\[
c^{(2)}_{\smallZ}=c^{(2)}\circ T^{(2)}_{\smallZ}.
\]
We find
\[
\Pi_\sigma c^{(2)}_{\smallZ}(\varphi\zotimes\psi)=
c^{(2)}_\sigma((\Pi_\sigma\varphi)\otimes(\Pi_\sigma\psi)).
\]

There is an essentially unique covariant representation of the DFR algebra
by self-adjoint coordinates \(q^\mu\); the commutators 
\(Q^{\mu\nu}=-i[q^\mu,q^\nu]\bar{\phantom I}\) strongly commute pairwise, and have
joint spectrum \(\Sigma\). By covariant we mean that there also is a strongly continuous unitary 
representation \(u\) of the Poincar\'e group fulfilling
\[
u(\varLambda,a)^{-1}q^\mu u(\varLambda,a)={\varLambda^\mu}_\nu q^\nu+a^\mu I.
\]
It follows that
\[
u(\varLambda,a)^{-1}Q^{\mu\nu} u(\varLambda,a)=
{\varLambda^\mu}_{\mu'}{\varLambda^\nu}_{\nu'} Q^{\mu'\nu'}.
\]
The quantisation of a generalised symbol \(\varphi=\varphi(\sigma;k)\) 
as described is given by
\[
\pi(\varphi)=\int dk\;\varphi(Q;k)e^{ikq},
\]
where the replacement of the dummy variable \(\sigma\) running in \(\Sigma\)
by \(Q\) must be understood in the sense of the
joint functional calculus of the operators
\(Q^{\mu\nu}\). \(\pi\) extends by continuity to a faithful, covariant 
representation of the dynamical system \((\mathcal E,\alpha)\), where
\[
u(L)\pi(T)u(L)^{-1}=\pi(\alpha(L)T),\quad T\in\mathcal E.
\]

This representation may be extended in a unique way to the multipliers
algebra \(M(\mathcal E)\); in this way, generalised symbols not vanishing
at infinity (as functions of \(\sigma\)) may also be quantised. This allows
to define
\[
W(f)=\pi(\check f)=\int dk\,e^{ikq}\check f(k).
\]

Due to the uniqueness (up to multiplicity and equivalence) 
of the covariant representation, we will often identify the Weyl operators
\(e^{ikq}\) and the twist operators \(e^{-(i/2)kQk}\) 
with elements of \(M(\mathcal E)\); and also with the corresponding 
generalised symbols. Under this {\itshape proviso}, we may write
\[
W_\sigma(f)=\Pi_\sigma W(f).
\]
Moreover,
\begin{equation}\label{eq:proj_cov_sigma}
W_\sigma(\gamma^{(1)}(\varLambda,a)f)=W_{\varLambda^{-1}\sigma{\varLambda^{-1}}^t}(\gamma^{(1)}(I,a)f),
\end{equation}
where we recall that \(\gamma^{(1)}(L)f(x)=f(L^{-1}x)\), and that \(f\) does not depend
on \(\sigma\).

\subsection{Twisted Covariance Recovered}\label{sec:eureka}

Let us define \(\mathscr T_\theta\) as the set of localisation states
\(\omega\) on the DFR algebra  
which are pure on the centre and concentrated on~\(\theta\), i.e.\ such that
\(\omega(f(Q))=f(\theta)\) for any \(f\in\mathcal C_0(\Sigma)\), where \(f(Q)\)
is the joint functional calculus of the pairwise strongly commuting operators \(Q^{\mu\nu}\);
in particular we have \(\omega(Q)=\theta\). This set is
evidently non invariant under the dual action of the Poincar\'e group; indeed a Poincar\'e
transformation \((\varLambda,a)\) maps \(\mathscr T_\theta\) onto 
\(\mathscr T_{\varLambda\theta\varLambda^t}\).

We now will show that the formalism of twisted covariance is equivalent to
constraining the fully covariant DFR model of quantum spacetime 
by means of the following {\itshape additional} assumption:
\begin{quote}{\bfseries \(\bm\theta\)-universality:} {\itshape there is class of equivalent
privileged 
observer; in the reference frame of a privileged observer, the only available
localisation states are precisely those in \(\mathscr{T}_\theta\); this non invariant
set transforms under the dual Poincar\'e action, when changing reference frame;} 
\end{quote}
we recall that~\(\theta\) is a universal datum fixed once and for all in the introduction. 

It is clear that the privileged observers are connected by Poincar\'e transformations in the 
stabiliser of~\(\theta\).

With the notations of section \ref{sec:dfr_recall}, the set of states available to the privileged observer is
\[
\mathscr T_\theta=\{\omega\circ\Pi_\theta:\omega\in\mathcal S(\mathcal K)\},
\]
where \(\mathcal S(\mathcal K)\) is the states space of \(\mathcal K\).

We set ourselves in a privileged reference frame. Since we only may test the algebra
with the states in \(\mathscr T_\theta\), we only can ``see'' the projections
\[
(\Pi_\theta\varphi)(\cdot)=\varphi(\theta,\cdot);
\]
it's like peeking through a narrow keyhole.
Here and below,  
the natural immersion \(\mathcal C_0(\Sigma,L^1)\subset\mathcal E\)
of the generalised symbols in the full algebra
is implicitly understood.

Now we perform a change in the reference frame: the new frame is connected to our privileged
one by the Poincar\'e transformation \(L=(\varLambda,a)\), 
and \(\theta'=\varLambda\theta\varLambda^t\).

In the full algebra, the section \(\varphi\) is mapped by the transformation to a 
new section \(\varphi'\) defined by
\[
\varphi'(\sigma,\cdot)=
e^{-ika}(\det\varLambda)\varphi(\varLambda^{-1}\sigma{\varLambda^{-1}}^t,
\varLambda^{-1}\cdot).
\]
The primed observer however 
would be bound by \(\theta\)-universality to project on the fibre
over \(\theta'\):
\[
(\Pi_{\theta'}\varphi')(\cdot)=\varphi'(\theta';\cdot)=
(\det\varLambda)e^{-ika}\varphi(\theta;\varLambda^{-1}\cdot);
\]
as expected, what she sees only depends on the original data at~\(\theta\).
Note that we may rewrite the above as
\[
\Pi_{\theta'}\varphi'=
\beta^{(1)}(L)(\Pi_{\theta}\varphi).
\]

Now we make the remark that both the observers we are considering, 
the privileged and 
unprivileged one, are not aware of the full structure of the algebra, 
since they cannot test
it. We may say that \(\theta\)-universality has turned the full structure of the algebra
into something somewhat metaphysical. The privileged observer, by making observations
in his own laboratory, cannot be expected to be so imaginative (or unwittingly 
complicated-minded) to devise all this structure under \(\theta\)-universality. 
He probably would develop instead
the algebra of the reduced commutation relations with matrix~\(\theta\); 
he would use functions depending on \(k\in\mathbb R^4\) 
only, not on \(\sigma\in\Sigma\), and define the twisted convolution \(\times_\theta\).
Analogously, the unprivileged observer, left alone, would not be aware of her unprivileged
status (which, after all, is only a convention: roles might well be exchanged) and
would define her own 
twisted convolution \(\times_{\theta'}\). They both would find the same algebra 
\(\mathcal K\) of compact operators, only with a different prescription for Weyl quantisation;
and they would be unaware of any problem until they would decide to compare their findings.

This situation is perfectly compatible with the remark that
\[
\Pi_\theta(\varphi\ztimes\psi)(k)=(\Pi_\theta\varphi)\times_\theta(\Pi_\theta\psi),
\]
in the frame of the privileged observer; and of course
\[
\Pi_{\theta'}(\varphi'\ztimes\psi')(k)=
(\Pi_{\theta'}\varphi')\times_{\theta'}(\Pi_{\theta'}\psi')
\]
in the unprivileged frame. 

Hence we completely reproduced the formalism of the reduced DFR model, 
which we already found equivalent to the formalism of twisted
covariance in subsection \ref{sec:wow}.

\subsection{Generalised Twisted Covariance}\label{sec:gen_cov}

DFR Weyl quantisation may be naturally 
generalised to functions taking values in some C*-algebra. We will discuss this in
some detail, in preparation of the discussion of third quantisation.

Let \(\mathcal F\) be any C*-algebra; then we may form the C*-algebra
\(\mathcal C_0(\mathbb R^4,\mathcal F)\) of continuous 
\(\mathcal F\)-valued functions vanishing at infinity, with pointwise 
multiplication:
\[
(fg)(x)=f(x)g(x),\quad f,g\in\mathcal C_0(\mathbb R^4,\mathcal F),
\]
where the product on the right hand side is taken in \(\mathcal F\);
the involution \(f\mapsto \bar f\) 
also is defined pointwise in terms of the involution
\(*\) of \(\mathcal F\): 
\[
\bar f(x)=f(x)^*;
\]
finally, the norm is 
\[
\|f\|=\sup\{\|f(x)\|_{\mathcal F}:x\in\mathbb R^4\}.
\]

The resulting algebra is commutative if and only if \(\mathcal F\) is 
commutative. In other words, it describes possibly 
noncommutative functions of 
a commutative space. This may be most easily seen if we consider the 
canonical isomorphism
\begin{equation}\label{eq:can_iso}
\mathcal C_0(\mathbb R^4,\mathcal F)\simeq
\mathcal C_0(\mathbb R^4)\otimes\mathcal F;
\end{equation}
the first factor is the localisation algebra; the second
factor is the range of the functions.

We may now formulate covariance: this requires that there is an action
\(\rho\) of the Poincar\'e group by automorphisms of \(\mathcal F\); 
we say that a certain function 
\(f\in\mathcal C_0(\mathbb R^4,\mathcal F)\) is covariant if it fulfils
\[
\rho(\varLambda,a)(f(x))=f(\varLambda^{-1}(x-a))
\quad (\varLambda,a)\in\mathscr P,x\in\mathbb R^4.
\]

The above may be rephrased on \(\mathcal C_0(\mathbb R^4)\otimes\mathcal F\), using
the canonical isomorphism \eqref{eq:can_iso}. With
\[
\gamma(L)(f)(x)=f(L^{-1}x)
\] 
on \(\mathcal C_0(\mathbb R^4)\), we say that 
\(f\in\mathcal C_0(\mathbb R^4)\otimes\mathcal F\) is covariant
if
\begin{equation}\label{eq:class_cov}
(\gamma(L)\otimes\id)(f)=(\id\otimes \rho(L))(f),\quad L=(\varLambda,a)
\in \mathscr P.
\end{equation}

The isomorphism \eqref{eq:can_iso} will be implicitly understood from now on.

Following our quantisation {\itshape ansatz}, we may replace the localisation algebra
\(\mathcal C_0(\mathbb R^4)\) by our new, quantised localisation algebra
\(\mathcal E\), namely  
\[
\mathcal C_0(\mathbb R^4,\mathcal F)\simeq
\mathcal C_0(\mathbb R^4)\otimes\mathcal F 
\leadsto
\mathcal E\otimes\mathcal F; 
\]
given the general structure of \(\mathcal E\),
the C*-tensor product is unique, and the resulting C*-algebra is isomorphic
to the trivial continuous field over \(\Sigma\) with standard fibre 
\(\mathcal K\otimes\mathcal F\).

This procedure of quantisation of the underlying geometry only affects the
first tensor factor; the algebraic structure of \(\mathcal F\) is
unaffected. We may regard \(\mathcal E\otimes\mathcal F\) as the algebra
of the continuous  functions of the non commutative spacetime 
which take values in \(\mathcal F\).

Recalling that the DFR algebra comes equipped with an action \(\alpha\)
of the Poincar\'e group, we may define an element \(X\in\mathcal E\otimes
\mathcal F\) as covariant if it fulfils
\[
(\alpha(L)\otimes\id)(X)=(\id\otimes \rho(L))(X),\quad L\in\mathscr P,
\]
by natural analogy with \eqref{eq:class_cov}.

Finally, DFR quantisation {\`a la Weyl} can be extended to 
\(\mathcal F\)-valued functions in the obvious way:
\[
\bm W(f)=\int dk\,e^{ikq}\otimes \check f(k),
\]
where both \(f\) and \(\hat f\) are in \(L^1(\mathbb R^4,\mathcal F)\).
Note that, with this definition
\[
\bm W=W\otimes\id:(L^1\cap \widehat{L^1})\otimes\mathcal F\rightarrow
M(\mathcal E)\otimes\mathcal F,
\]
where \(W\) is the ordinary DFR quantisation {\itshape \`a la Weyl}.

Note that the DFR quantisation intertwines the
actions of the Poincar\'e group on the classical and quantised function algebra:
\[
\bm W(\gamma(L)f)=(\alpha(L)\otimes\id)(\bm W(f)),
\quad L\in\mathscr P,
\]
so that \(\bm W(f)\) is covariant if and only if \(f\) is covariant.

It may happen (and it happens, indeed) that \(f\) only is covariant under the
restricted Poincar\'e group; in which case the above condition of covariance must
be restricted accordingly. 

The Weyl calculus can be developed as usual; now to close it we need
generalised symbols with values in \(Z\otimes\mathcal F\); with the
usual identification \(Z=\mathcal C_b(\Sigma)\) of the centre \(Z\)
of the multipliers algebra \(M(\mathcal E)\), we may think of a symbol
as of a function of \(\Sigma\times\mathbb R^4\), taking values
in \(\mathcal F\). Hence
\[
\bm W(f)\bm W(g)=\bm W(f\star g),
\] 
where 
\[
(f\star g)\!\check{\phantom{I}}(\sigma,k)=(\check f\times \check g)(\sigma,k)=
\int dh\check f(h)\check g(h-k)e^{-\frac{i}{2}h\sigma k}.
\]
Also the action on generalised symbols is the usual one.

We may define the projection 
\[
\bm \Pi_\sigma=\Pi_\sigma\otimes\id:\mathcal E\otimes\mathcal F\rightarrow 
\mathcal K\otimes\mathcal F
\]
onto the fibre over \(\sigma\),
and reproduce straightforwardly 
the discussion of the preceding section in terms of the reduced Weyl
quantisation
\[
\bm W_\sigma=W_\sigma\otimes\id=\bm\Pi_\sigma \bm W.
\]

Let us again restrict ourselves to Schwartz symbols, for the sake of simplicity: we
denote by \(\mathscr S^{(n)}_{\mathcal F}\) the set of
Schwartz \(\mathcal F^{n\otimes}\)-valued symbols of \(n\) variables, 
and we implicitly understand the isomorphism with 
\(\mathscr S(\mathbb R^{4n})\otimes\mathcal F^{n\otimes}\) (as a l.c.s). 

We denote as usual by \(m^{(n)}\) the \(n\)-fold pointwise product 
\(
m^{(n)}:\mathscr S(\mathbb R^{4n})\rightarrow \mathscr S(\mathbb R^4)\)
of complex valued symbols, and by \(\mathcal M^{(n)}:\mathcal F^{n\otimes}\rightarrow\mathcal F\)
the product \(\mathcal M^{(n)}(F_1\otimes\dotsm\otimes F_n)=F_1\dotsm F_n\) 
in the C*-algebra \(\mathcal F\); we then define the product
of generalised symbols
as 
\[
M^{(n)}=m^{(n)}\otimes \mathcal M^{(n)}:
\mathscr S^{(n)}_{\mathcal F} \rightarrow \mathscr S^{(1)}_{\mathcal F}
\]
We now again fix~\(\theta\) in a given reference frame; the twisted product is
\[
M^{(n)}_\theta=m^{(n)}_\theta\otimes M^{(n)}=
(m^{(n)}(F^{(n)}_\theta\cdot)\otimes \mathcal M^{(n)}.
\]
Of course, in momentum space we take \(C^{(n)}=c^{(n)}\otimes\mathcal M^{(n)}\)
and \(C^{(n)}_\theta= c^{(n)}_\theta\otimes\mathcal M^{(n)}\).

The ordinary and twisted Poincar\'e actions are
\begin{gather*}
\Gamma^{(n)}(L)=\gamma^{(n)}(L)\otimes \id_{\mathcal F^{n\otimes}}
\end{gather*}
and the twisted action is
\[
\Gamma^{(n)}_\theta(L)=\gamma_\theta^{(n)}(L)\otimes \id_{\mathcal F^{n\otimes}};
\]
once again \(\Gamma^{(1)}=\Gamma^{(1)}_\theta\). 
Twisted covariance then reads
\[
M^{(n)}_\theta\circ\Gamma^{(n)}_\theta(L)=\Gamma^{(1)}(L)\circ M^{(n)}_\theta.
\]

Twisting covariance may be seen as adding correction terms to the coproduct, in order
to compensate the choice of forcing~\(\theta\) to be constant. If we restrict ourselves to
covariant symbols, i.e.\ symbols fulfilling \eqref{eq:class_cov}, we may 
obtain an equivalent result by twisting the coproduct of the action \(\rho\) on \(\mathcal F\)
instead of the action \(\gamma\) on \(\mathcal C_0(\mathbb R^4)\). Note however that
the resulting twisted action \(\Rho^{(n)}_\theta\) only does the expected job
in restriction to classically covariant symbols.

Let us define 
\[
\Rho^{(n)}(L)=\id_{\mathscr S^{4n}}\otimes \rho(L)^{n\otimes}
\]
By definition, a covariant symbol \(f\in\mathscr S^{n}_{\mathcal F}\) fulfils
\[
\Gamma^{(n)}(L)f=\Rho^{(n)}(L)f.
\]
We seek for a modification \(\Rho^{(n)}_\theta(L)\) such that, for any covariant symbol \(f\),
\[
\Gamma^{(n)}_\theta(L)f=\Rho_\theta^{(n)}(L)f.
\]
With \(\theta'=\varLambda_L\theta\varLambda_L^t\), the right hand side 
of the above may be rewritten as
\(
F_{-\theta}^{(n)}F_{\theta'}^{(n)}\gamma(L)^{n\otimes}\otimes \id_\mathcal F^{n\otimes}f\) which in 
turn, using the covariance of the symbol, equals
\(F_{-\theta}^{(n)}F_{\theta'}^{(n)}\otimes \rho(L)^{n\otimes}f\); we have thus the solution
\[
\Rho_\theta^{(n)}(L)=F_{-\theta}^{(n)}F_{\theta'}^{(n)}\otimes\rho(L)^{n\otimes},
\]
or
\[
\tilde\Rho^{(n)}(L)=T_{-\theta}^{(n)}T_{\theta'}^{(n)}\otimes\rho(L)^{n\otimes}
\]
in momentum space. We may observe that the idea of swapping the twist of the coproduct
from the first to the second tensor factor of 
\(\mathcal S(\mathbb R^{4n})\otimes\mathcal F^{n\otimes}\) is an optical illusion; 
the twist only acts on the first factor, as it is made clear by the different forms it takes
according to whether we are in position or momentum space 
(which only makes sense in the first factor).

\section{Third Quantisation}\label{sec_main:fields}

In this section we will show that, even in the reduced DFR model (i.e.\ under
\(\theta\)-universality), third
quantised fields according to the DFR prescription {\itshape \`a la Weyl} are 
covariant with respect to the undeformed action of the special Poincar\'e group
\(\mathscr P_+^\uparrow\), if \(\theta\) is properly treated as a tensor.

In addition, we will describe the results of \cite{lg_1,buchholz,gandalf-harald}
on two purposes: 1) to clarify their relations with the models discussed here,
and 2) because they provide a convenient framework to discuss the covariance properties
of the so called twisted CCR introduced in \cite{bal_1,bal_2}. We will show that
\(\theta\)-universality is either not assumed or unnecessary, in the above mentioned
approaches.

\subsection{DFR Quantisation}\label{dfr_q_fields}

The third quantisation 
\[
\phi(q)=\bm W(\phi)=\int dk\,e^{ikq}\otimes \check \phi(k)
\]
of the free massive boson field was first proposed in \cite{dfr}.
It can be morally understood as the DFR quantisation of a ``function'' 
\(\phi=\phi(x)\) of the classical spacetime, taking values 
``in'' the field algebra \(\mathcal F\). Up to carefully rephrasing
everything in terms of tempered distributions and affiliation, we 
are essentially in the situation described in 
subsection \ref{sec:gen_cov}.
We refrain from spelling the details, which are standard.

Let \(U\) be the usual strongly continuous unitary representation 
of the restricted Poincar\'e group \(\mathscr P_+^\uparrow\) on the Fock space. The free field
\(\phi\) is covariant, namely it fulfils
\[
\rho(L)\phi(x)=\phi(L^{-1}x),\quad L\in\mathscr P_+^\uparrow,
\]
where \(\rho(L)\) is the adjoint action of \(U(L)\):
\[
\rho(L)\phi(x)=U(L)\phi(x)U(L)^{-1}.
\]

Correspondingly, the third quantised field is covariant, too:
\[
(\alpha(L)\otimes\id)(\bm W(\phi))=(\id\otimes \rho(L))(\bm W(\phi)),\quad L\in
\mathscr P_+^\uparrow.
\] 

Now, we remark that \((\id\otimes \rho(L))(\bm W(\phi))=\bm W(\rho(L)\phi)\);
by this and \eqref{eq:proj_cov_sigma}, the above implies
\[
\bm W_{\sigma}(\gamma^{(1)}(L)\phi)=\bm W_{\varLambda\sigma\varLambda^t}(\rho(L)\phi);
\]
where \(L=(\varLambda,a)\in\mathscr P_+^\uparrow\), and \(\gamma^{(1)}(L)\phi(x)=\phi(L^{-1}x)\).

It follows from the above remarks that
\[
\rho(L)(\phi\star_\sigma\phi)=
(\rho(L)\phi)\star_{\sigma'}(\rho(L)\phi),\quad \sigma'=\varLambda\sigma\varLambda^t;
\]
hence the formalism of twisted covariance may be equivalently
applied if we assume \(\theta\)-universality, 
and the fields are twisted covariant with respect to the usual (undeformed)
representation of the restricted Poincar\'e group on the Fock space if one
keeps~\(\theta\) invariant in all reference frames.

Since free fields are covariant, we might apply the ideas of subsection \ref{sec:gen_cov}
and realise the formalism of twisted covariance by twisting  
the coproduct associated to the representation of the restricted Poincar\'e group 
on the Fock space instead.

Though possible, we feel that this step has more disadvantages than advantages. First of
all, as discussed in full detail in subsection \ref{sec:gen_cov}, twisting the coproduct 
on the Fock side induces an action which is wrong by definition when applied to {\itshape non}
covariant  fields; 
this would lead to systematic (and probably uncontrollable)
errors when dealing e.g.\ with the perturbative theory of an interactive field with
infrared cut-off (which breaks covariance until removed); and even at a formal level
without infrared cutoff, in all known approaches to perturbation theory (which, as of today,
all break covariance under Lorentz boosts; see e.g.\ \cite{dfr,ultraviolet}). Secondly,
it conveys the not undebatable feeling that, in this particular class of models, 
noncommutativity of spacetime can be transferred into the
definition of the Fock space; indeed, as we made explicit in subsection \ref{sec:gen_cov},
twists always act on the localisation algebra, even if we let them be artificially
carried by the twisted coproduct on the Fock space.

\subsection{Wedge Locality and Warped Convolutions}\label{sec:bgl}
In preparation of the next subsection, we shortly review the results of 
\cite{lg_1,buchholz,gandalf-harald}.

Let
\[
\mathcal W_0=\{x:x^1>|x^0|\}\subset\mathbb R^4
\]
be the standard wedge (sometimes called the right wedge by analogy with theories in
1+1 dimensions). In \cite{lg_1} the class of antisymmetric matrices \(\sigma_0\in\Sigma\) 
fulfilling the following conditions has been characterised:
\begin{itemize}
\item{(i)} if \(L=(\varLambda,a)\in\mathscr L_+^\uparrow\) is such that 
\(L\mathcal W_0\subset\mathcal W_0\), then \(\varLambda\sigma_0\varLambda^t=\sigma_0\);
\item{(ii)} if \(L=(\varLambda,a)\in\mathscr L_+^\uparrow\) is such that 
\(L\mathcal W_0\subset\mathcal W_0'\), then \(\varLambda\sigma_0\varLambda^t=-\sigma_0\);
\item{(iii)} \(\sigma_0V_+=\mathcal W_0\);
\end{itemize}
where \(V_+\) is the future timelike cone, and the prime
indicates the causal complement if applied to regions of spacetime (or the commutant
if applied to sets of bounded operators).
The characterisation is obtained by observing that each \(\sigma_0\) as above 
and \(W_0\) must have the same stabiliser in \(\mathscr L^\uparrow_+\). In what follows
we fix a choice of \(\sigma_0\) as above.

Let \([\mathcal W]\) denote the equivalence class of wedges containing \(\mathcal W\), where 
two wedges are said equivalent if they can be obtained from each other by translations;
moreover, let \([\mathcal W]_0\) be the unique element of that class whose edge contains 
the origin.
Next, choose a continuous map
\(\sigma\mapsto\varLambda_\sigma\) fulfilling 
\(\varLambda_{\sigma}\sigma_0{\varLambda_{\sigma}}^t=\sigma\) (which exists, but of course is not
unique; see \cite{dfr}), and define the 
map \([\mathcal W]\mapsto\sigma([\mathcal W])\) 
by requiring that \(\varLambda_{\sigma([\mathcal W])}\mathcal W_0=[\mathcal W]_0\).

Motivated by the results of \cite{lg_1}, an abstract construction (called warped convolution) was 
introduced, leading to the definition of a nonlocal, wedge-local net
\(\mathcal W\mapsto\mathcal F(\mathcal W)\) of W*-algebras,
which are obtained by deformation (warped convolution) of an existing local theory; 
for each wedge \(\mathcal W\), 
the parameter of the deformation is precisely \(\sigma([\mathcal W])\). If the
undeformed theory is covariant, isotonic and fulfils the Reeh-Schlieder property with respect
to \(\Omega\), so does the deformed theory w.r.t.\ the same representation of 
\(\mathscr P_+^\uparrow\). Moreover, if the undeformed theory is local, the deformed theory
is wedge-local:
\[
\mathcal F(\mathcal W')\subset\mathcal F(\mathcal W)'.
\]
Note that
the resulting net does neither depend on the initial choice of \(\sigma_0\),
nor on the choice of the map \(\sigma\mapsto\varLambda_\sigma\).

To investigate the relations of the above setting with our results, we take the point of
view of \cite{gandalf-harald}, where the authors generalised their previous work also
in the light of \cite{buchholz}. For our purposes it will be sufficient to cast ourselves
in a simplified setting, where there is one only massive neutral spin 0 free field; \(\mathcal H\)
is the Fock space, and \(\Omega\) the vacuum vector. With the pairing
\[
\langle \bm W(\phi),f\rangle=\int dx\; \phi(q+x)f(x), 
\quad f\in\mathcal S(\mathbb R^4),
\]
the third quantised field algebra is the smallest W*-algebra 
\(\bm{\mathcal F}\) to which all the operators  \(\langle\bm W_\sigma(\phi),f\rangle\),
\(f\in\mathscr S^{(1)}\), are affiliated to\footnote
{In \cite{gandalf-harald} the polynomial field algebra is considered instead, which allows
for more general Wightman fields to encompass the results of \cite{buchholz}; 
here we concentrate on the free field, in which case
the present formulation is equivalent to that of \cite{gandalf-harald}.}. 
For each 
\(\sigma\in\Sigma\), we make a choice
\(\omega_\sigma\) of a pure state on \(\mathcal K\); we then consider the 
GNS representation \((\bm\pi^{\omega_\sigma},\mathcal H^{\omega_\sigma},\Omega^{\omega_\sigma})\) 
of \(\bm{\mathcal F}\) with respect
to the state \((\omega_\sigma\circ\Pi_\sigma)\otimes(\Omega,\cdot\Omega)\restriction_{\bm{\mathcal F}}\).
It is extended as usual to the unbounded operators 
affiliated to \({\bm{\mathcal F}}\), so
that we may define the fields
\[
\phi^{\omega_\sigma}(f)=\bm\pi^{\omega_\sigma}(\langle \bm W(\phi),f\rangle).
\] 
on \(\mathcal H^{\omega_\sigma}\). In \cite{gandalf-harald} it is shown that
there is a family  \(\{\phi^{\sigma}:\sigma\in\Sigma\}\) of non local fields
on the Fock space \(\mathcal H\), and invertible linear isometries 
\(V^{\omega_\sigma}:\mathcal H^{\omega_\sigma}\rightarrow\mathcal H\),
fulfilling the following properties:
\begin{gather*}
V^{\omega_\sigma}\Omega^{\omega_\sigma}=\Omega,\\
\phi^\sigma (f)V^{\omega_\sigma}=V^{\omega_\sigma}\phi^{\omega_\sigma}(f),
\quad f\in\mathscr S(\mathbb R^4),\\
U(L)\phi^\sigma(f)U(L^{-1})=
\phi^{\varLambda\sigma\varLambda^t}(\gamma^{(1)}(L^{-1})f),\quad L\in\mathscr P_+^\uparrow; 
\end{gather*}
in particular, the covariant family \(\{\phi^\sigma:\sigma\in\Sigma\}\) does not depend
on the particular choice of \(\omega_\sigma\) for each \(\sigma\), provided it is of the
required type.

Let us now define \(\mathcal F(\mathcal W)\) as the smallest W*-algebra to which all fields
of the form \(\phi^{\sigma([\mathcal W])}(f)\), \(\text{supp}\,f\subset \mathcal W\), 
are affiliated. According to \cite{gandalf-harald,buchholz}, the 
net \(\mathcal W\mapsto\mathcal F(\mathcal W)\) is precisely the same
wedge-local, non local net as the one obtained by means of warped convolution.

The approach of \cite{lg_1,buchholz,gandalf-harald} is not based on 
assumptions of the kind of \(\theta\)-universality, 
but provides instead a novel tool for constructing a 
fully covariant, wedge-local, nonlocal theory on ordinary Minkowski spacetime.

However,
the construction is driven uniquely by the geometry of wedges in the (classical) spacetime,
and it is not clear how could it be interpreted 
as a (possibly effective) theory on quantised spacetime. 
We will discuss this and related questions  in the next subsection.

\subsection{Fibrewise Twisted CCR}\label{sec:ftccr}

The fields \(\phi^\sigma\) described in the preceding subsection can be explicitly 
constructed by twisting the tensor product of the Borchers-Uhlmann algebra 
\cite{gandalf-harald}. It is not clear, however, that within the original 
interpretation there is any relation with commutation relations 
among the coordinates,
other than initial motivation. Indeed, the twists of the tensor products are 
different (in general) for different wedges in the same reference frame, so that a specific
twist cannot be attached to the coordinates of the frame itself. All the deformed fields 
are available to each observer, who for every wedge builds the corresponding field algebra
by appropriately picking the corresponding field in the covariant family \(\{\phi^\sigma\}\). 

Of course, one might well take a completely different view, and make an arbitrary
choice of a pair \((\theta,\mathcal O)\) of a matrix \(\theta\in\Sigma\) and  
of a (privileged) Lorentz observer \(\mathcal O\); in terms of this one might 
postulate that the field theory in that particular frame is described by the field 
\(\phi^\theta\). We are precisely in the setting of \(\theta\)-universality.

In this way, one may reproduce the formalism of twisted commutation relations 
developed in \cite{bal_1,bal_2}
\begin{align*}
a^{\sigma}(p_1)a^{\sigma}(p_2)=&e^{-ip_1\sigma p_2}a^{\sigma}(p_2)a^{\sigma}(p_1),\\
a^{\sigma}(p_1){a^{\sigma}}^\dagger(p_2)=&e^{ip_1\sigma p_2}
{a^{\sigma}}^\dagger(p_2)a^{\sigma}(p_1)+\\&+p^0\delta^{(3)}(\vec p_1-\vec p_2),
\end{align*}
where \(p_1,p_2\) are on the forward mass shell. With these relations,
\[
\phi^\theta(x)=\int dp\,\delta(p^2-m^2)\theta(p^0)
\left(e^{ipx}{a^\theta}^\dagger(p)+e^{-ipx}a^\theta(p)\right).
\]

The above relations can be realised by defining 
\[
a^\sigma(p)=e^{\frac{i}{2}p\sigma P}a(p),\quad
{a^\sigma}^\dagger(p)=e^{\frac{i}{2}p\sigma P}a^\dagger(p),
\]
where \(p\) is on shell and \(a,a^\dagger\) are the usual (undeformed) creations and 
annihilations on the Fock space of the (undeformed) free theory.

We may use these remark to show that even the machinery of twisted commutation relations
does not rely on \(\theta\)-universality.

Indeed, disregarding the original motivations for the construction of the fields 
\(\phi^\sigma\) we may use them as building blocks for a new representation of the fields
\(\bm W(\phi)\) described in subsection \ref{dfr_q_fields}.

Consider in fact the fields
\[
\phi^Z(f)=\int^\oplus d\varLambda\, \phi^{\varLambda\sigma_0\varLambda^t}(f) 
\]
as operators on 
\[
\mathcal H^Z=\int^\oplus d\varLambda\, \mathcal H\simeq 
L^2(\mathscr L,d\varLambda)\otimes\mathcal H,
\]
where of course \(d\varLambda\) is the Haar measure of the full Lorentz 
group.

Define on the dense subspace
of measurable vector fields \(\Psi:\varLambda\mapsto \mathcal H\) the unitary representation
\[
(U^Z(L)\Psi)(M)=U(L)\Psi(\varLambda^{-1}M),\quad L=(\varLambda,a)\in\mathscr P_+^\uparrow;
\]
By construction, this gives a covariant field
\[
U^Z(L)\phi^Z(f)U^Z(L)^{-1}=\phi^Z(\gamma^{(1)}(L)f).
\]
The map \(\pi^Z(\langle\bm W(\phi),f\rangle)= \phi^Z(f)\) induces
a representation of \(\bm{\mathcal F}\) on \(\mathcal H^Z\), which we still
denote by \(\pi^Z\). Moreover,
with 
\[
\bm\alpha=\alpha\otimes\id\restriction_{\bm{\mathcal F}}=
\id\otimes\rho(L)\restriction_{\bm{\mathcal F}},
\]  
then \((\pi^Z,U^Z)\) is a covariant, faithful representation of the
W*-dynamical system \((\bm{\mathcal F},\bm\alpha)\).

Of course,
\begin{gather*}
a^Z(p)=\int^\oplus d\varLambda\,e^{\frac{i}{2}p(\varLambda\sigma\varLambda^t)P}a(p),\\
{a^Z}^\dagger(p)=\int^\oplus d\varLambda\,e^{-\frac{i}{2}p(\varLambda\sigma\varLambda^t)P}a^\dagger(p)
\end{gather*}
fulfil fully (undeformed) covariant fibrewise twisted commutation relations, which
we will analyse elsewhere.

\section{Conclusions}\label{sec:conclusions}
We have shown that the formalism of twisted covariance may be described equivalently
by superposing a non invariant constrain (\(\theta\)-universality) 
on otherwise admissible localisation states 
of the DFR model of quantum spacetime. Concerning quantum field theory on quantum spacetime, 
we have shown that the formalism of 
twisted tensor product and twisted CCR does not require \(\theta\)-universality to be
assumed, and can be understood fibrewise, in a fully covariant way.

This raises some 
strong concerns about statements on possible observable effects of 
\(\theta\)-universality.

In other words, \(\theta\)-universality 
does not seem to be a
necessary assumption in any of the approaches considered here: it appears as unnecessary
both when quantising the spacetime alone, and when attempting quantum field theory on it. Note also that, even in the framework of twisted covariance, partial
indications of the survival of the undeformed Lorentz group 
already appeared in the literature, at the cost of distinguishing 
so called particle
transformations from observer transformations \cite{bgg,gb,gb1}; 
in addition, there were already indications that the twisted structure 
does not seem to allow for accommodating more field content than the 
reduced DFR model \cite{zahn}.

As a matter of fact, \(\theta\)-universality implies
a fundamental breakdown of the relativity principle: notwithstanding that, as we saw,
form--covariance may be restored, still it is possible to classify
the observers according to the particular \(\theta'=\varLambda\theta\varLambda^t\) 
which is attached to their
Lorentz frame. Although 
covariance might well be replaced by a more fundamental concept at Planck scale, 
yet we should not forget the intrinsic
limits of the class of models we are discussing here, which are conceived so
to represent a somewhat 
``semiclassical'' quantisation of the flat Minkowski spacetime, and which we may 
expect to allow for describing 
at best a limited class of processes. 
In this framework, \(\theta\)-universality would have observable consequences
also in the large scale limit, which would contradict the excellent experimental
fittings for special relativity in its range of validity\footnote{Sergio Doplicher 
publicly advocated this view on many occasions in the last fifteen years.}.

Even putting aside the above somewhat philosophical remarks and
landing on very concrete grounds, 
we have shown here that working with the fully covariant
DFR model is equivalent to drop \(\theta\)-universality. Hence in all approaches considered 
here \(\theta\)-universality
was not at all forced upon us by the interpretation, 
but was instead an optical illusion due to the
particular formalism adopted. 

Indeed, the DFR model is fully covariant as far as free fields are concerned.
Every attempt to define interactions on this model entailed the breakdown
of covariance under Lorentz boosts 
at some level \cite{dfr,ultraviolet,quasiplanar}. 
This may eventually be traced to the fundamental, unsolved  problem  of
devising an adequate noncommutative replacement for the concept of locality.

In the author's opinion, strong physical motivations (or experimental indications, whenever they
will become available) should be provided to justify \(\theta\)-universality within the expected
range of validity of this particular class of models; since, otherwise, 
a fully covariant formalism is available for the localisation algebra, which cannot be rejected for free.

{\footnotesize \subsubsection*{Acknowledgements}
I am deeply indebted with Sergio Doplicher for many enlightening 
and enjoyable conversations on this topic, and his comments on a preliminary
version of this work. Gandalf Lechner patiently explained me some aspects of his
joint results with Harald Grosse. I also gratefully acknowledge Ludwik Dabrowski for
his support and constructive comments.
Last but non least, I thank Claudia for her smile, and the blue sky.}

\appendix

\section{Many Variables}
\label{app:many_var}

Functions \(f(x_1,\dotsc,x_n)\) of many variables may be studied under two point of view,
which both are useful and allows for the formulation of different problems. Already classically,
we may think of \(x_j=x_1+a_j\) as translations of one point of coordinates \(x_1\), 
or as independent degrees of freedom. 

These two approaches can be reproduced on the quantised spacetime, where however 
(at least in the approach we are discussing here) quantisation only affects the coordinates,
while translations remain classical.

\subsection{Translations of a Single Event}

We first consider translations of one single localisation event: then one
may wish to give meaning to objects of the form \(f(q+a_1,q+a_2,\dotsc,q+a_r)\).
We let ourselves be guided by the special case \(f=f_1\otimes\dotsm\otimes f_r\), where
the notations themselves lead us to the natural definition
\[
(f_1\otimes\dotsm\otimes f_r)(q+a_1,q+a_2,\dotsc,q+a_r)=f_1(q+a_1)\dotsm f_r(q+a_r),
\]
from which we immediately derive the general definition 
\[
f(q+a_1,q+a_2,\dotsc,q+a_r)=m^{(r)}(F^{(r)}_\theta f_{\bar a})(q),
\]
where \(f_{\bar a}(x_1,\dotsc,x_r)=f(x_1-a_1,\dotsc,x_r-a_r)\).

This definition was for example considered in \cite{dfr}, where it was shown that the commutator
of an optimally localised field with its own translate by \(a\) falls off exponentially in 
any spacelike direction as a function of the Euclidean length \(|a^2|\) of the 
displacement \(a\), when evaluated on
an optimally localised (i.e.\ coherent) state. 

An apparently third party choice for the coordinates of many events has been proposed recently
by \cite{fw}. There, quantum coordinates \(\hat x_i^\mu\) are considered,
which fulfil
\begin{equation}\label{puzzano}
[\hat x_j^\mu,\hat x_k^\nu]=i\theta^{\mu\nu},\quad j,k=1,2,\dotsc,n.
\end{equation}
(no \(\delta_{jk}\)), 
namely the many localisation events are not considered independent. 

At first sight, one could object that relations of this kind would introduce Planck scale
correlations between events separated by no matter how large distances 
(even at cosmic scales), which sounds at least implausible.
As the author themselves observed, however, the differences of such 
coordinates are central (``classical variables''). 
It follows that the relations \eqref{puzzano}
only have trivial irreducible representations: we rephrase Remark 2
of \cite[Sect.\ 2]{fw} as 
\begin{lemma}
Let \(\hat x_j^\mu\), \(j=1,\dotsc,n\), \(\mu=0,\dotsc,3\), 
self-adjoint operators 
fulfilling \eqref{puzzano} strongly (i.e. in Weyl form) and irreducibly. 
Then there are 
\(n-1\) real 4-vectors \(a_2,\dotsc,a_n\) such that
\[
\hat x_j^\mu=\hat x^\mu_1+a^\mu_j,\quad j=2,\dotsc, n.
\]
\end{lemma}

\noindent{\itshape Proof.} \([\hat x_1^\mu,(\hat x_j-\hat x_1)^\nu]=0\) 
strongly, hence by 
Schur's lemma \(\hat x_j^\nu-\hat x_1^\nu=a_j^\nu\).\qed

In other words, the relations \eqref{puzzano} are equivalent to consider 
the coordinates of one single event, together with its classical translations.
Such coordinates, then, may be useful to study self--correlations of a single
localisation event (as in \cite{dfr}). However, their interpretation as 
coordinates of many events would contradict the folk lore about localisation 
at short distances. Indeed, under such an interpretation we would be forced to 
allow for the separation between independent events to be observed with 
arbitrary precision.

\subsection{Many Independent Events: Symbol Calculus and Twi\-sted Covariance}

The other natural possibility\footnote{A variant of this choice could be to take different~\(\theta\)'s in different tensor factors; we shall discuss it briefly in the next subsection.} 
is to consider independent localisation events of coordinates
\[
q_{\theta j}^\mu=I\otimes\dotsm \otimes I\otimes q_\theta^\mu\otimes I
\otimes\dotsm
\otimes I
\quad\textrm{(\(r\) factors, \(q_\theta^\mu\) in the \(j^{\text{th}}\) slot)}.
\]
Of course these coordinates fulfil
\[
[q^\mu_{\theta j},q_{\theta k}^\nu]=i\delta_{jk}\theta^{\mu\nu}.
\]

The universal enveloping C*-algebra of these relations is again the algebra of
compact operators \(\mathcal K\) on the separable infinite dimensional Hilbert space, and
the Weyl quantisation 
\begin{align*}
W^{(r)}_\theta(f)&=
\int dk_1\dotsm dk_r\,\check f(k_1,\dotsc,k_r)e^{ik_1q_\theta}\otimes\dotsm\otimes 
e^{ik_rq_\theta}=\\&=
\int dk_1\dotsm dk_r\,\check f(k_1,\dotsc,k_r)e^{i\sum_jk_jq_{\theta j}}
\end{align*}
induces an isomorphism 
\(\mathcal K\simeq\mathcal K^{n\otimes}\) via 
\[
W_\theta^{(r)}(f_1\otimes\dotsm\otimes f_r)=
W_\theta(f_1)\otimes\dotsm\otimes W_\theta(f_r).
\]

There is an induced twisted product of functions of \(r\) variables which
is the natural product in the tensor product algebra of symbols:
\[
(f_1\otimes\dotsm\otimes f_r){\star_\theta}(g_1\otimes\dotsm\otimes g_r)
=(f_1{\star_\theta} g_1)\otimes\dotsm\otimes(f_r{\star_\theta} g_r)
\]
and equivalently a tensor product of twisted convolutions in momentum space.

The above product of functions may be equivalently described as a twisted product:
defining 
\[
R^{(n,r)}\bigotimes_{k=1}^n\bigotimes_{j=1}^rf^k_j=\bigotimes_{j=1}^r\bigotimes_{k=1}^nf^k_j
\]
and the multitwist
\[
F_\theta^{(n,r)}=(\underbrace{F_\theta^{(n)}\otimes\dotsm \otimes F_\theta^{(n)}}_{\textrm{\(r\) factors}})
R^{(n,r)},
\]
then the twisted product of \(n\) symbols of \(r\) variables is
\[
m_\theta^{(n,r)}(f\otimes g)=m^{(n,r)}(F_\theta^{(n,r)}f\otimes g),
\]
where
\[
m^{(n,r)}\left(\bigotimes_{j=1}^nf_j\right)(x_1,\dotsc,x_r)=
\left(\prod_{j=1}^nf_j\right)(x_1,\dotsc,x_r)
\]
is the ordinary pointwise product.

Moreover, the twisted action of the Poincar\'e group becomes
\[
\gamma^{(n,r)}_\theta(L)={F^{(n,r)}_\theta}^{-1}\gamma^{(n,r)}(L) F^{(n,r)},
\]
where the untwisted action is
\begin{gather*}
(\gamma^{(n,r)}(L)f_1\otimes\dotsm\otimes f_n)(x^1_1,\dotsc,x^1_r,\dotsc,x^n_1,\dotsc,x^n_r)=\\
(\gamma^{(1,r)}(L)f_1)(x^1_1,\dotsc,x^1_r)\dotsm(\gamma^{(1,r)}(L)f_n)(x^n_1,\dotsc,x^n_r)=\\
=f_1(L^{-1}x^1_1,\dotsc,L^{-1}x^1_r)\dotsm f_n(L^{-1}x^n_1,\dotsc,L^{-1}x^n_r).
\end{gather*}
The corresponding twisted coproduct is
\begin{align*}
\Delta^{(2,r)}_\theta[X]&={F^{(2,r)}_\theta}^{-1}\left(
\Delta[X]\otimes\Delta[I]+\Delta[I]\otimes\Delta[X]
\right)F^{(2,r)}_\theta=\\&=R^{(2,r)}\left(\Delta_\theta[X]\otimes\Delta_\theta[I]+
\Delta_\theta[I]\otimes \Delta_\theta[X]\right)R^{(2,r)}.
\end{align*}
With these notations, twisted covariance reads
\[
m^{(n,r)}_\theta(\gamma^{(n,r)}_\theta(L)f_1\otimes \dotsm\otimes f_n)=
\gamma^{(1,r)}(L)m^{(n,r)}_\theta (f_1\otimes \dotsm\otimes f_n).
\]
The proof that
\[
m^{(n,r)}_\theta(\gamma^{(n,r)}_\theta(L)f_1\otimes \dotsm\otimes f_n)=
m^{(n,r)}_{\theta'}\big((\gamma^{(1,r)}(L)f_1)\otimes\dotsm\otimes(\gamma^{(1,r)}(L) f_n)\big)
\]
with \({\theta'}^{\mu\nu}={\varLambda^\mu}_\mu'{\varLambda^\nu}_\nu'{\theta}^{\mu'\nu'}\)
is the obvious adaptation of the same argument for \(r=1\).

\subsection{Many Independent Events in the Fully Covariant  DFR Algebra}

For the sake of completeness, we provide a short account of the fully covariant approach
to many independent events.

When taking into account the full DFR algebra, there are two inequivalent definitions
of coordinates of many independent events.

One possibility is to take
\[
q_j^{\mu}=I^{(j-1)\otimes}\otimes q^\mu\otimes I^{(r-j)\otimes},\quad j=1,\dotsc,r,
\]
so that, with \(Q^{\mu\nu}_j=-i[q^{\mu}_j,q^\nu_j]\!\bar{\phantom{I}}\),
\begin{equation}\label{eq:DFR_full_tensor}
[q_j^\mu,q_k^\nu]=i\delta_{jk}Q^{\mu\nu}_j,\quad [Q_j^{\mu\nu},Q_k^{\mu\nu}]=0
\end{equation}
strongly, where each of the tensors \(Q_1,\dotsc,Q_r\) fulfils the DFR constrain.
 
These relations have an essentially unique covariant representation, and the resulting universal
enveloping C*-algebra \(\mathcal E^{r\otimes}\)
is isomorphic to \(\mathcal C_0(\Sigma^r,\mathcal K^{r\otimes})\simeq
\mathcal C_0(\Sigma^r,\mathcal K)\); the 
corresponding symbols are then functions 
of \(\Sigma^n\times \mathbb R^{4n}\). 

Taking the above definition, it would be possible to recover the discussion of many variables
of the preceding  subsection assuming \(\theta\)-universality, 
by taking as admissible localisation states all those which are pure
on the centre of \(\mathcal E^{r\otimes}\) 
and concentrated on \((\theta,\theta,\dotsc,\theta)\in\Sigma^r\). 

The above immediately
suggests that one might consider as well different~\(\theta\)'s for the coordinates of 
different events, which would amount to select localisation states pure
on the centre and concentrated on \((\theta_1,\dotsc,\theta_r)\) with \(\theta_j\neq\theta_k\) 
(possibly). The development of the
corresponding formalism is straightforward, but we refrain from spelling the details also in view
of our fundamental criticism of \(\theta\)-universality.

A different choice is to replace the relations \eqref{eq:DFR_full_tensor}  with
\begin{equation}\label{eq:DFR_Z_tensor}
[q_j^\mu,q_k^\nu]=i\delta_{jk}\mathcal Q^{\mu\nu}
\end{equation}
where the commutators \(\mathcal Q\) of independent coordinates (not the coordinates themselves!)
are identified; namely we divide the algebra of the relations \eqref{eq:DFR_full_tensor} 
by the differences \(Q_i-Q_j\). In other words, we consider 
the coordinates 
\[
q_j^{\mu}=I^{(j-1)\zotimes}\zotimes q^\mu\zotimes I^{(r-j)\zotimes},\quad j=1,\dotsc,r,
\]
where \(\zotimes\) is the tensor product of \(Z\)-moduli over the centre \(Z\) of the multipliers
algebra \(M(\mathcal E)\), so that
\[
[q_1^\mu,q_1^\nu]=\dotsb=[q_r^\mu,q_r^\nu]=i\mathcal Q^{\mu\nu}
\]
and \(\mathcal Q\) fulfils the DFR constrains. The resulting algebra \(\mathcal E^{r\zotimes}\)
is isomorphic to \(\mathcal E\); the symbols associated to Weyl quantisation are functions
of \(\Sigma\times\mathbb R^{4r}\).

Also with this choice we may derive the formalism of many variables of the preceding
subsection, assuming \(\theta\)-universality. 

This choice appears more natural than taking the ordinary tensor product, in that it amounts
to treat noncommutativity (encoded in the manifold \(\Sigma\)) as background--independent data.

With this choice, the differences of coordinates cannot be made arbitrarily 
small, but are bound to limitations at the same scale than the coordinates themselves. These remarks were first used in \cite{ultraviolet}, 
where a new notion
of Wick product for the \(\phi^n\) self-interaction on quantum spacetime
was constructed; 
the corresponding unitary S-matrix was found free of ultraviolet divergences, 
as an effect of the regularisation induced by spacetime quantisation.

\subsection{Wightman Functions}
According to the preceding discussion, there are two natural definitions
of Wightman functions in this context, which for simplicity we discuss 
``at fixed \(\theta\)''. 
Let \(\phi(x)\) be a local (second quantised) 
field, and \(\phi(q)\) its third quantisation. The first 
possibility could be to naively set
\[
\mathscr W_\theta(q_1,\dotsc,q_n)=(\Omega,\phi(q_1)\dotsm\phi(q_n)\Omega),
\]
where \(\Omega\) is the vacuum of the local theory.
However this definition would not sense any noncommutativity; indeed, taking 
\(n\) localisation states \(\omega_1,\omega_2,\cdots,\omega_n\), we might
evaluate
\[
\left\langle\omega_1\otimes\dotsm\otimes\omega_n,
\mathscr W_\theta(q_1,\dotsc,q_n)\right\rangle=
\mathscr W(f_1\otimes\dotsm\otimes f_n)
\]
where \(\mathscr W\) is the Wightman function of the initial local theory, and
\(\hat f_i(k)=\omega_i(e^{ikq})\). No twists show up. 

Apparently more promising would be to take instead
\[
\mathscr W_\theta(q;x_1,\dotsc,x_n)=
(\Omega,\phi(q+x_1)\dotsm\phi(q+x_n)\Omega),
\]
depending on the classical parameters \(x_j\).
Unfortunately, by smearing this object with a test function 
\(f=f(x_1,\dotsc,x_n)\) and evaluating the resulting object with 
a (sufficiently regular) localisation state \(\omega\), we would get
\[
\left\langle\omega,\mathscr W_\theta(q;f)\right\rangle=
\mathscr W(K_\omega f),
\]
where again \(\mathscr W\) is the local Wightman function, 
and\footnote{It is a general fact that, for generic test functions \(f,g\),
\begin{gather*}
\int f(q+x_1,\dotsc,q+x_n)g(x_1,\dotsc,x_n)dx_1\dotsm dx_n=\\
=\int f(x_1,\dotsc,x_n)g(q-x_1,\dotsc,q-x_n)dx_1\dotsm dx_n,
\end{gather*}
namely it is legitimate to perform the change of integration variables
\(x_j\rightarrow q-x_j\); this classical recipe survives spacetime 
quantisation because (as it may be easily checked) everything can be done 
while preserving the relative order of Weyl operators.
}
\[
(K_\omega f)(x_1,\dotsc,x_n)=\langle\omega,f(x_1-q,\dotsc,x_n-q)\rangle
\]
defines a nonlocal operator acting on test functions. In other words, also with 
this definition, nonlocality is encoded in the localisation algebra and
there is no interplay with the fields.

Indeed, this is precisely what should be expected. We are facing an essentially
perturbative approach, where the local field is the zero order, and 
noncommutativity shows up as higher order perturbation terms. No interesting
non commutativity should be expected from the spacetime quantisation of a local 
field, since spacetime quantisation is kinematical, and the initial field 
content is local.

\end{document}